\documentclass{iau} 
\usepackage{graphicx}

\title[Principal component analysis of geomagnetic activity] 
{Principal component analysis of geomagnetic activity: New information on solar wind \\}

\author[Kalevi Mursula \& Lauri Holappa]   
{Kalevi Mursula,
\and Lauri Holappa}

\affiliation{ReSoLVE Centre of Excellence, Space Climate Research Unit,  University of Oulu, Finland\\ email: {\tt kalevi.mursula@oulu.fi} \\[\affilskip]}

\pubyear{2017}
\volume{335}  
\jname{Space Weather of the Heliosphere: Processes and Forecasts}
\editors{C. Foullon \& O. Malandraki}
\begin{document}

\maketitle


\begin{abstract}

We use the principal component analysis (PCA) to study geomagnetic activity at annual resolution using a network of 26 magnetic stations in 1966-2015, and an extended network of 40 stations in 1980-2015. The first principal component (PC1) describes the long-term evolution of global geomagnetic activity, and has an excellent correlation with indices like the Kp/Ap index. The two networks give identical results for PC1. The second principal component (PC2) is highly correlated with the annual percentage of high-speed streams (HSS). The extended network has a slightly higher sensitivity to HSSs. We verify the non-trivial latitudinal distribution of the second empirical orthogonal function (EOF2). We find that the amplitude of the 22-year variation of geomagnetic activity has a closely similar latitudinal distribution as EOF2. This verifies that the 22-year variation of geomagnetic activity is related to HSSs. The most likely cause is the Russell-McPherron mechanism. 

\keywords{Space climate, Geomagnetic activity, High-speed solar wind, Principal component analysis}
\end{abstract}


\firstsection 

\section{Local geomagnetic activity in a global network of stations}

Geomagnetic activity is a measure of the effect of the solar wind and the accompanied heliospheric magnetic field to the Earth's magnetic field.
This interaction drives various electric current systems in the magnetosphere and ionosphere that are monitored by ground-based magnetometers as geomagnetic activity.
The two main solar wind structures driving geomagnetic activity are the coronal mass ejections (CME) and high speed solar wind streams together with the related corotating interaction regions (CIRs).
CMEs and HSS/CIRs are responsible for the most severe types of geomagnetic activity, the substorms and storms.

We extend our earlier analysis [\cite[Holappa et al., 2014]{Holappa_2014}], where we used data from a network of 26 ground-based magnetic stations from 1966 to 2009, both temporally, including data from 1966 until 2015 and spatially by using an extended network of 40 stations from 1980 to 2015.
Both networks include stations from a large range of latitudes and longitudes in the northern hemisphere. 
The criteria for choosing the observatories were the continuity, length and quality of data, and the global coverage of stations. 
The list of the original 26 stations is given in Table 1 of \cite[Holappa et al., 2014]{Holappa_2014}.
The additional stations of the extended network are 
Guam, Honolulu, Hurbanovo, Furstenfeldbruck, Boulder, Novosibirsk, Saint Johns, Newport, Ottawa, Yellowknife, Barrow, Hornsund, Cambridge Bay, and Resolute Bay. 
The data for these magnetic stations were retrieved from the World Data Center of Edinburgh, UK. 

\begin{figure}
	\centering
	\includegraphics[width=0.64\linewidth]{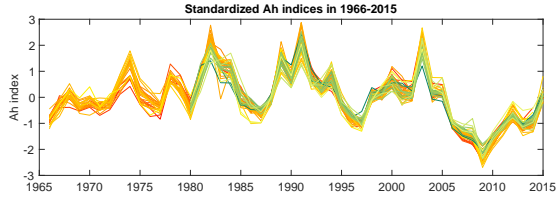}
\caption{Annual means of $A_h$ indices of local geomagnetic activity for 26 stations in 1966-2015 and for 14 stations in 1980-2015.} 
\label{Ah_timeseries}
\end{figure}

Figure \ref{Ah_timeseries} shows the annual $A_h$ indices [\cite[Mursula and Martini, 2007]{Mursula_Martini_2007}] for the 26 stations from 1966 to 2015 and the additional 14 stations from 1980 to 2015.
$A_h$ index is a measure of local geomagnetic activity, which follows the K-index method but uses hourly digital magnetic data, and therefore can be calculated straightforwardly for all stations whose magnetic field measurements are given 
as digital hourly values. 
We have standardized (remove mean; normalize by standard deviation) the annual $A_h$ means for each station in order to remove the latitudinal variation of geomagnetic activity and to have all stations on an equal footing for PCA.
Figure \ref{Ah_timeseries} shows a very uniform temporal evolution of geomagnetic activity, with roughly equal relative solar cycle variation in all stations.
In addition, one can see the recent decadal decline of geomagnetic activity, with annual means from 2006 until 2014 remaining below zero. 
Only in 2015 the level of geomagnetic activity reached the long-term mean (zero).
This shows the exceptional quietness of the near-Earth space in recent years.

\section{Principal component analysis}

\begin{figure}[h!]
	\begin{center}
		\includegraphics[width=0.95\linewidth]{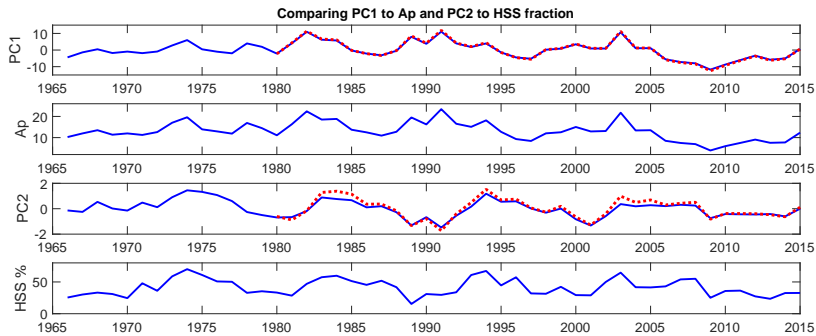} 
\caption{[top] PC1 for 26 stations in 1966-2015 (solid line) and for 40 stations in 1980-2015 (dotted line) [second] Global Ap index of global geomagnetic activity [third] PC2 for 26 stations in 1966-2015 (solid line) and for 40 stations in 1980-2015 (dotted line) [bottom] Annual percentages of HSSs in solar wind.}
\label{PC1_PC2_Ap_HSS}
	\end{center}
\end{figure}

We have made the PCA for the 26 stations in 1966-2015 and, separately, for the 40 (26+14) stations in 1980-2015.
Figure \ref{PC1_PC2_Ap_HSS} shows PC1 and PC2 of these two networks in panels 1 and 3, respectively.
In both cases PC1, which gives a measure of global geomagnetic activity, takes some 96\% of the total variance in  annual means, and the two PC1s can be hardly distinguished from each other in Fig. \ref{PC1_PC2_Ap_HSS}.
PC1 is highly correlated with the Ap index, which can, based on this comparison, be verified as a reliable measure of global geomagnetic activity at annual scale.

PC2 explains only some 2\% of total variance in annual means in both cases.
The large fraction of
PC1 should, however, not mask the important information included in PC2.
The two PC2s depict small differences between each other around 1984, 1994 and 2003.
These are the times of most intense HSSs, as shown in Fig. \ref{PC1_PC2_Ap_HSS} (bottom panel). 
Correlation coefficient between the extended network and HSS percentage is 0.86, while it is 0.83 for the 26 stations during the same time interval in 1980-2015 (and 0.81 over the whole time 1966-2015).
This difference is the virtue of using a larger station network, which takes a better account of the HSS effect to geomagnetic activity in different regions of the world.
Anyway, in both cases, the HSSs explain a very large fraction of the variability of PC2 (73\% and 70\%, respectively), demonstrating their crucial importance for PC2.

\begin{figure}[h!]
	\begin{center}
		\includegraphics[width=0.85\linewidth]{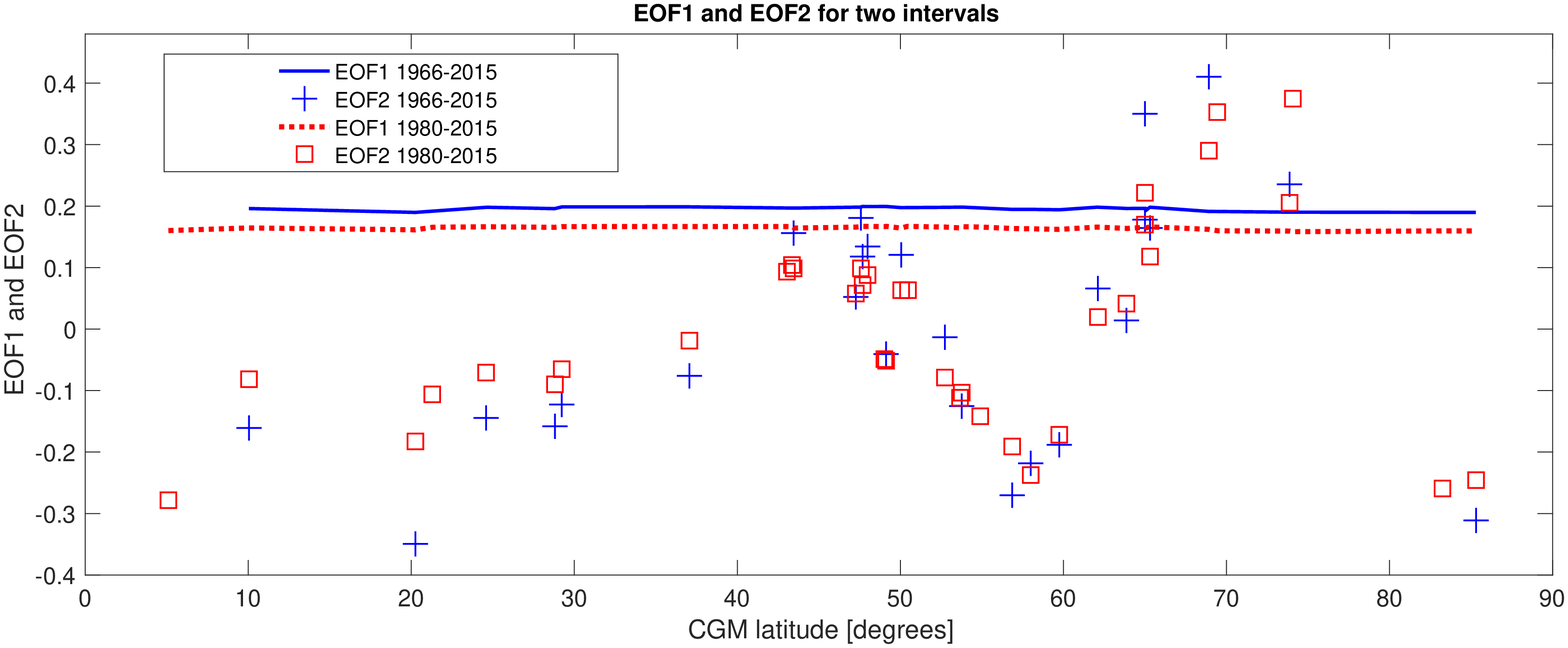} 
\caption{EOF1 (lines) and EOF2 (pluses and squares) for 26 stations in 1966-2015 (solid line) and for 40 stations in 1980-2015 (dotted line) as a function of corrected geomagnetic latitude.}
\label{EOF1_EOF2}
	\end{center}
\end{figure}

EOFs (scalings) of the first two principal components are depicted in Figure \ref{EOF1_EOF2}.
EOF1s are almost straight lines, indicating that the fraction of PC1 is roughly the same for all stations of a network.
(EOF1 of the extended network is smaller because of the larger number stations).
The latitudinal distribution of EOF2 has a maximum at the auroral oval, a local minimum at subauroral latitudes at about $55-60^{\circ}$ and another maximum at mid-latitudes of about $40^{\circ}$-$50^{\circ}$.
This distribution is due to the difference in the average location and intensity of substorms related to CMEs and HSSs [\cite[Holappa et al., 2014]{Holappa_2014}].
At auroral latitudes the effect of HSSs to geomagnetic activity is largest, leading to the peak in EOF2.
Towards lower latitudes, the effect of HSSs decreases but the decrease is not uniform with latitude.
Rather, the relative effect of CMEs is largest at sub-auroral latitudes, leading to the local minimum of EOF2 at these latitudes.

\section{Solar magnetic cycle at different latitudes}

Two successive 11-year solar activity cycles affect to the Earth's magnetic field slightly differently, producing a 22-year variation of geomagnetic activity.
This is related to the variation of solar polarity during the 22-year magnetic cycle, also called the Hale cycle.
A mechanism which naturally produces a polarity dependent response in alternating cycles is the Russell-McPherron (R-M) mechanism [\cite[Russell and McPherron, 1973]{Russell_McPherron_1973}].
The horizontal component of the heliospheric magnetic field produces, because the tilt of the Earth's rotation axis with respect to solar equator, a component which can enhance or suppress geomagnetic activity depending on solar polarity.  
During positive polarity times the R-M effect enhances geomagnetic activity (in spring and fall), and reduces that in negative polarity times [\cite[Cliver et al., 1996]{Cliver_etal_1996}].
 
\begin{figure}[h!]
	\begin{center}
		\includegraphics[width=0.75\linewidth]{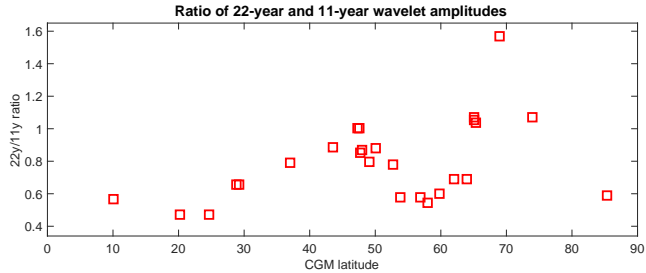} 
\caption{Ratio of the wavelet amplitudes related to the 22-year Hale cycle and to the 11-year Schwabe cycle for 26 stations in 1966-2015 as a function of corrected geomagnetic latitude.}
\label{Wavelet_ratio2211}
	\end{center}
\end{figure}

We have calculated the wavelet spectrograms of daily $A_h$ indices for the 26 stations using Morlet wavelet. 
Figure \ref{Wavelet_ratio2211} depicts the ratio between the wavelet amplitudes of 22-year and 11-year variations as a function of corrected geomagnetic latitude.
The latitudinal variation of this ratio is strikingly similar to that of the EOF2.
Accordingly, the Hale cycle in geomagnetic activity is much stronger at auroral and even at mid-latitude stations than in sub-auroral stations.
As explained above, PC2 describes the fraction of HSSs in solar wind, and the latitudinal variation of EOF2 shows where the HSS effect is relatively larger or smaller compared to CMEs.
Therefore, the similarity of the latitudinal variation of the 22-year amplitude and EOF2 strongly suggests that the Hale cycle of geomagnetic activity is due to effects related to high-speed streams.
A natural explanation is the R-M mechanism since the polarity of HSSs faithfully follows the dominant polarity in the source hemisphere.
We have also verified that the Hale ratio maximizes during the minimum times of solar cycle, when the solar polarity is most clearly ordered with heliographic latitude (not shown here).
At these times the  R-M mechanism enhances (reduces) geomagnetic activity during positive (negative, resp.) polarity minima.

\section{Conclusions}

We have used here the principal component analysis to study long-term geomagnetic activity.
PC1 describes the global geomagnetic activity, and is quite insensitive to the number of stations used in PCA.
It correlates well with the traditional Kp/Ap index, thus verifying the validity of Kp/Ap as a measure of global geomagnetic activity.
PC2 is highly correlated with the annual percentage of high-speed streams.
We found very similar results for PC2 from the two networks, although the extended network has a slightly better sensitivity to high-speed streams.
We verified the interesting, non-trivial latitudinal distribution of the EOF2.
We also found that the amplitude of the 22-year (Hale cycle) variation of geomagnetic activity has a closely similar latitudinal distribution as EOF2. 
This demonstrates that the Hale cycle of geomagnetic activity is due to the effect of HSSs.
The most likely cause is the Russell-McPherron mechanism. 


\section*{Acknowledgements}
We acknowledge the financial support by the Academy of Finland to the ReSoLVE Centre of Excellence (project no. 272157).
Magnetometer data were obtained from the World Data Center for Geomagnetism, Edinburgh (\texttt{http://www.wdc.bgs.ac.uk/}), and solar wind data from the OMNI database (\texttt{http://omniweb.gsfc.nasa.gov/}).
We appreciate the continuous work made at the observatories around the world.



%

\end{document}